\documentclass{iaus}
\usepackage{graphicx}

\def\BP{Ballesteros-Paredes}
\def\Eg{E_{\rm g}}
\def\Ek{E_{\rm k}}
\def\Eth{E_{\rm th}}
\def\gtsima{$\; \buildrel > \over \sim \;$}    
\def\gtrsim{\lower.5ex\hbox{\gtsima}}           
\def\kms{{\rm km~s}^{-1}}
\def\ltsima{$\; \buildrel < \over \sim \;$}    
\def\ld{\lambda_{\rm d}}
\def\lesssim{\lower.5ex\hbox{\ltsima}}           

\def\ls{\lambda_{\rm s}}
\def\MJ{M_{\rm J}}
\def\MJe{M_{\rm J,eff}}
\def\Ms{{\cal M}_{\rm s}}
\def\pcc{{\rm ~cm}^{-3}}
\def\psc{{\rm ~cm}^{-2}}
\def\sun{{\rm sun}}
\def\tc{\tau_{\rm c}}

\def\tff{\tau_{\rm ff}}

\def\VS{V\'azquez-Semadeni}

\title[Molecular Cloud Turbulence and the Star Formation Efiiciency] 
{Molecular Cloud Turbulence And The Star Formation Efficiency: Enlarging
the Scope}

\author[\VS]   
{Enrique \VS$^1$}

\affiliation{$^1$Centro de Radioastronom\'\i a y Astrof\'\i sica, UNAM,
Campus Morelia, P.O. Box 3-72, Morelia, Michoac\'an, 58088, M\'exico}

\pubyear{2006}
\volume{xxx}  
\date{?? and in revised form ??}
\jname{Proceedings Title IAU Symposium}
\editors{A.C. Editor, B.D. Editor \& C.E. Editor, eds.}
\begin{document}

\maketitle

\begin{abstract}
We summarize recent numerical results on the control of the star
formation efficiency (SFE), addressing the effects of turbulence and
the magnetic field strength. In closed-box numerical simulations, the
effect of the turbulent Mach number $\Ms$ depends on whether the
turbulence is driven or decaying: In driven regimes, increasing $\Ms$
with all other parameters fixed decreases the SFE, while in decaying
regimes the converse is true. The efficiencies in non-magnetic cases for
realistic Mach numbers $\Ms \sim 10$ are somewhat too high compared to
observed values. Including the magnetic field can bring the SFE down to
levels consistent with observations, but the intensity of the magnetic
field necessary to accomplish this depends again on whether the
turbulence is driven or decaying. In this kind of simulations, a
lifetime of the molecular cloud (MC) needs to be assumed, being
typically a few free-fall times. Further progress requires determining
the true nature of the turbulence driving and the lifetimes of the
clouds. Simulations of MC formation by large-scale compressions in the
warm neutral medium (WNM) show that the generation of the clouds'
initial turbulence is built into the accumulation process that forms
them, and that the turbulence is driven for as long as accumulation
process lasts, producing realistic velocity dispersions and also thermal
pressures in excess of the mean WNM value. In simulations including
self-gravity, but neglecting the magnetic field and stellar energy
feedback, the clouds never reach an equilibrium state, but rather evolve
secularly, increasing their mass and gravitational energy until they
engage in generalized gravitational collapse. However, local collapse
events begin midways through this process, and produce enough stellar
objetcs to disperse the cloud or at least halt its collapse before the
latter is completed. Simulations of this kind including the missing physical
ingredients should contribute to a final resolution of the MC lifetime
and the origin of the low SFE problems.

\end{abstract}

\firstsection 
\section{Introduction} \label{sec:intro}

Molecular clouds (MCs) are the densest regions in the interstellar
medium (ISM) and also the site of all present-day star formation
in the Galaxy. They are known to have masses much larger than their
thermal Jeans mass, a fact that led \cite{GK74} to propose that the
clouds should be in a state of generalized gravitational
collapse. However, \cite{ZP74} readily noted that this would imply that
the MCs should be forming stars at very high rates
($\sim 30 M_\sun$ yr$^{-1}$) if all of their mass were to be
transformed into stars in roughly one free-fall time $\tff$, while the
observed rates are much lower ($\sim 5 M_\sun$ yr$^{-1}$; see, e.g.,
\cite{SP04}), suggesting that the star formation efficiency (SFE) is reduced
by some mechanism. The observed SFE ranges from a few percent when whole
giant molecular complexes are considered (e.g., \cite{Myers_etal86}), to
10--30\% in cluster-forming cores (e.g., \cite{LL03}).

The SFE in MCs, defined as the fraction of the clouds' mass that
finally makes it into a star during their lifetime, can be
simply written as SFE = ${\rm SFR} \times
\Delta \tc$, where SFR is the star formation rate, and $\Delta \tc$ is
the cloud lifetime. Thus, a low SFE can be obtained through
either a small $\Delta \tc$ or a low SFR. Currently, there is an ongoing
debate within the community on whether the cloud lifetimes are long, but
the SFR is small (or even zero) over a major fraction of the cloud
lifetime (\cite[e.g.]{PS00}; \cite{PS02};
\cite{TM04}; \cite{MTK06}; \cite{TKM06}), or else the lifetimes are short,
but the SFRs are relatively large (e.g., \cite{BHV99}; \cite{KHM00};
\cite[Hartmann, \BP\ \& Bergin 2001, herafter HBB01]{HBB01}; \cite{Hart03};
\cite{BBB03}; \cite{BB06}; \cite{BH06}; \cite{VS_etal06b}). 

MCs are also known to have supersonic linewidths, which have been
attributed to turbulent motions (e.g., \cite{ZE74}; \cite{Larson81};
\cite{Blitz93}). Turbulent flows are characterized by a scaling of the
typical velocity difference $\Delta v$ across points separated by a
distance $\ell$ that scales as $\Delta v \propto \ell^\alpha$, with
$\alpha >0$ (e.g., \cite{Lesieur90}), implying that the largest velocity
differences occur at the largest spatial scales of a given cloud or
clump (\cite{Larson81}). Thus, turbulence is expected to have a {\it
dual} role in the dynamics of MCs (e.g., \cite{VP99}; \cite{MK04};
\cite{BKMV06}): On the one hand, with respect to regions of size $L$,
supersonic compressive turbulent modes of size $\ell > L$ will act
mainly as pistons that can form a density peak (a ``cloud'', ``clump''
or ``core'') out of those regions. The timescale for clump formation is
essentially the turbulent crossing time across scale $\ell$. Since the
compressions are supersonic, this is typically shorter than the
free-fall or sound-crossing times. On the other hand, turbulent modes
with $\ell < L$ will provide support against the self-gravity of a clump
of size $L$.

In this paper we review recent numerical results concerning the effect of the
molecular cloud turbulence and the magnetic field on the regulation of
the SFE, and discuss how the resolution of certain issues, such as the
determination of the most appropriate set of parameters requires
studying the formation of the clouds themselves. This review extends
the one presented earlier by \cite{VS05}.

\section{Effect of the driving scale and turbulent Mach number on the SFE}
\label{sec:Eff_M}

\begin{figure}
\scalebox{0.5}{\centerline{\includegraphics[height=4in,width=4in]{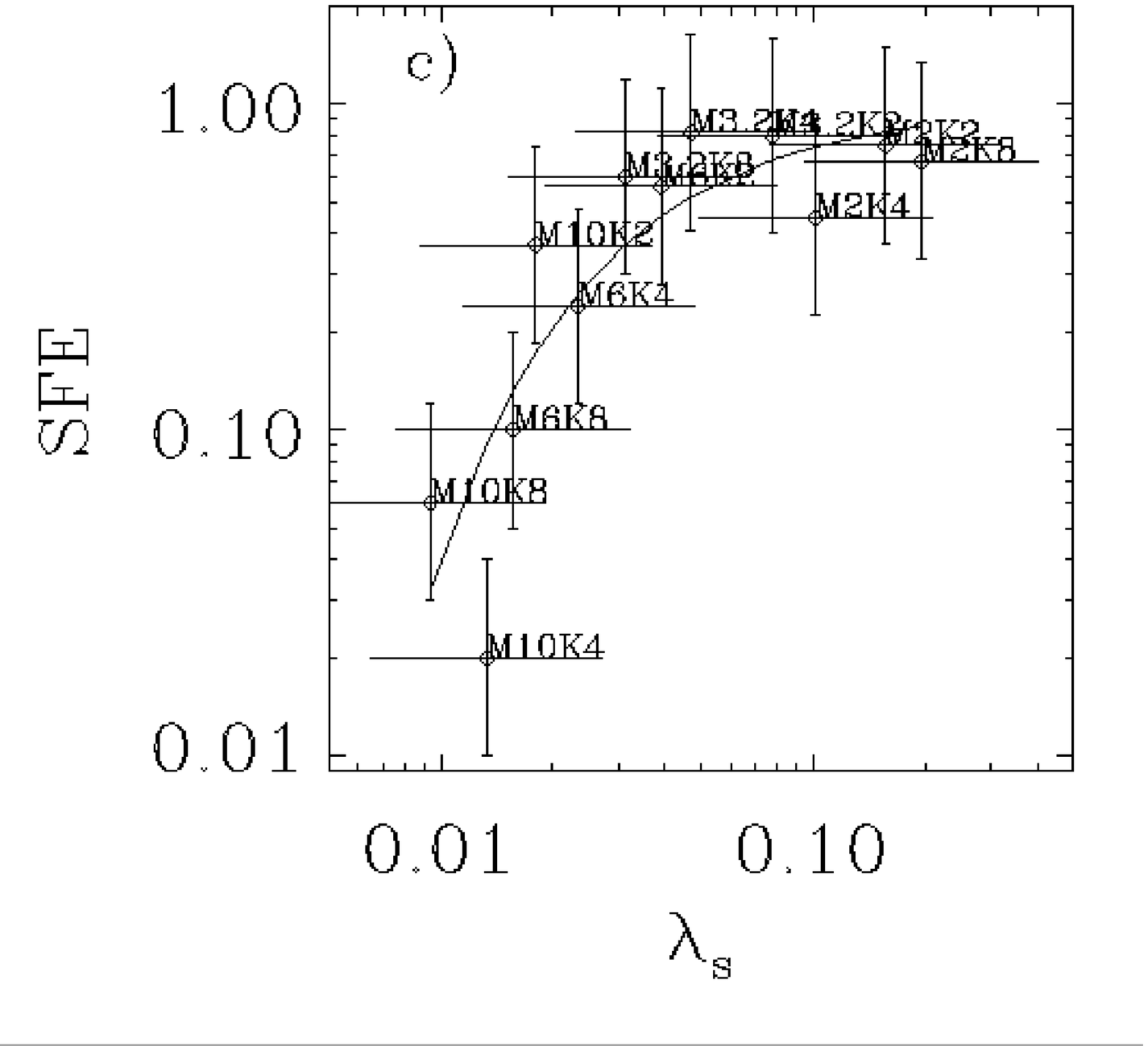}}}
\scalebox{0.5}{\centerline{\includegraphics[height=6in,width=6in]{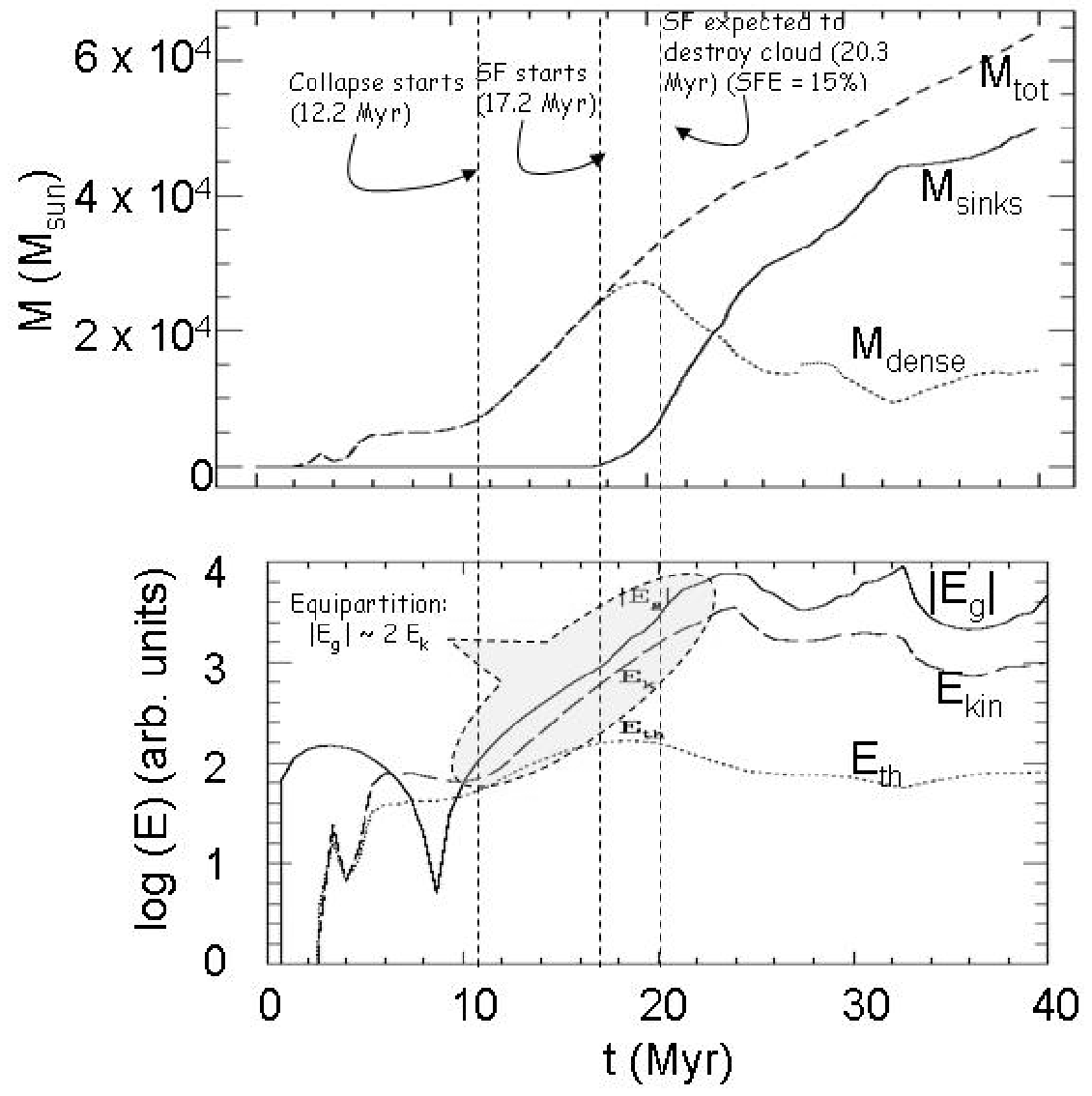}}}
  \caption{\emph{Left panel:} Star formation efficiency SFE vs.\ sonic scale
$\ls$ for runs with various rms Mach numbers (M) and turbulence driving
wavenumbers K, indicated for each point (from \cite[\VS\ et
al. 2003]{VBK03}). \emph{Right panel:} Evolution of the dense gas
mass, $M_{\rm dense}$, mass in stellar objects $M_{\rm sinks}$, and total
mass $M=M_{\rm dense} + M_{\rm sinks}$ (\emph{top}) and the
gravitational, kinetic and thermal energies (resp.\ $\Eg$, $\Ek$ and
$\Eth$) for a simulation of colliding WNM streams in the presence of
thermal bistability and self-gravity (\emph{bottom}, from \cite[\VS\ et al.\
2006b]{VS_etal06b}).\label{fig:SFE_non-MHD}}
\end{figure}

Results from numerical simulations in the recent past have shown that
the effect of the rms Mach number of the turbulence $\Ms$ on the SFE
depends on whether the turbulence is driven or decaying. In continuously
driven regimes in closed boxes, with periodic boundary conditions and a
fixed total mass, 
\cite[Klessen et al.\ (2000)]{KHM00} 
showed that the SFE decreases systematically as either the driving scale
 of the turbulence $\ld$ is
decreased, or the turbulent Mach number $\Ms$ is increased, and
\cite[V\'azquez-Semadeni, Ballesteros-Paredes \& Klessen (2003)]{VBK03}
subsequently showed that the dependence on $\Ms$ and $\ld$ could be
combined into the dependence with one single parameter, the \emph{sonic
scale} $\ls$ of the turbulence. This is the scale at which the typical
turbulent velocity fluctuation (which decreases with scale) equals the
sound speed, and is related to $\Ms$ and $\ld$ by $\ls \approx \ld
\Ms^{-1/\alpha}$, where $\alpha$ is the exponent in the
velocity dispersion-size relation (cf., \S \ref{sec:intro}). 
At a fixed number of Jeans masses, reducing $\ls$
leads to a reduction of the fraction of the
total mass in scales smaller than $\ls$, and the SFE is
expected to decrease. Indeed, \cite[V\'azquez-Semadeni et al.\
(2003)]{VBK03} were able to empirically fit a functional dependence of
the form SFE $\propto \exp{(-\lambda_0/\ls)}$, with $\lambda_0 \sim
0.11$ pc in the simulations they studied (fig.\ \ref{fig:SFE_non-MHD},
\emph{left panel}). 
From the above relation, this translates into SFE $\approx \exp{(-\lambda_0 
\Ms^{1/\alpha}/\ld)}$, implying that, at fixed $\ld$, $\ls$ decreases
with increasing $\Ms$ in driven regimes. This can be understood in terms
of the net effect of the turbulent velocity fluctuations, which on the
one hand produce larger-amplitude density fluctuations, but on the other
increase the effective ``sound'' speed in the flow, giving the net
result that the effective Jeans mass $\MJe$ scales with the rms Mach
number as
\begin{equation}
\MJe \propto \Ms^2 \label{eq:MJ_Ms}
\end{equation}
(\cite{MK04}, sec.\ IV.G). At larger $\MJe$ it becomes increasingly
difficult to collect a core more massive than this mass that can proceed to
collapse. 
 
In contrast, in decaying regimes, the SFE appears to \emph{increase}
with the rms Mach number of the initial velocity fluctuations
(\cite{NL05}). This can be understood because, in this regime,
the initial velocity fluctuations can still perform the same fast
clump-forming role as in driven regimes. However, at later times, the
global decay of the turbulence implies that its supporting action is
gradually lost, and $\MJe$ decreases, as indicated
by eq.\ (\ref{eq:MJ_Ms}) for $\Ms$ decreasing over time. 

In either case, the efficiencies obtained in non-magnetic numerical
simulations still appear larger than observational values. For
example, \cite[\VS\ et al.\ (2003)]{VBK03} reported an SFE $\sim
30$\% in a driven simulation with $\Ms = 10$ and a mass $M = 64 \MJ$
($=1860 M_\sun$), where $\MJ$ is the thermal Jeans mass,
while a two-dimensional decaying simulation with \emph{initial} $\Ms =
10$ and $M = 100 \MJ$ reported by \cite[Nakamura \& Li (2005)]{NL05}
reached SFE $\sim 60$\%. Note also that the latter simualtion actually
had already decayed to $\Ms \sim 2$--3 by the time it was forming stars,
a value that appears too low compared with typical turbulent Mach numbers
observed in clouds (e.g., \cite{Blitz93}).


\section{Effect of the magnetic field strength on the SFE}
\label{sec:Eff_B}

The magnetic field is an important physical ingredient of interstellar
dynamics, and may contribute towards further reducing the SFE obtained
in simulations to levels more consistent with observations, even in
magnetically supercritical regimes.

In the magnetic case, a fundamental control parameter is the
mass-to-magnetic flux ratio $\mu$ (in units of the critical value for
magnetic support against collapse). Under ideal MHD conditions,
supercritical cases ($\mu >1$) can undergo gravitational collapse, while
subcritical cases ($\mu < 1$) are unconditionally supported against
it. In this case, collapse can only occur if a Lagrangian fluid
parcel loses some of its magnetic flux through some dissipative or
diffusive process, such as ambipolar diffusion (AD; e.g.,
\cite{Mes_Spit56}). 


Numerical simulations show that, in magnetically supercritical simulations,
collapse is in general delayed with respect to the non-magnetic case
(\cite{OGS99}; \cite{HMK01}; \cite{VKSB05}; \cite{NL05}). Recently, the
SFE has been measured in simulations of 3D, driven, supercritical
simulations of ideal MHD (\cite{VKSB05}) and decaying, 2D simulations
including AD (\cite{NL05}). Realistic values of the SFE at the level of
whole clouds (SFE $\sim$ a few percent) required moderately subcritical
regimes in the decaying cases, but only moderately supercritical regimes in
the driven simulations, evidencing again the distinction between driven
and decaying regimes.  Stronger fields are needed in decaying conditions
to compensate for the systematic loss of turbulent support.


In any case, both types of studies show that the SFE is reduced by the
presence of a magnetic field even in supercritical regimes, with a
tendency to greater reductions at larger mean field strengths. This
suggests that the effect of the magnetic field on attenuating the SFE
may be gradual rather than dychotomic, as was the case of the
distinction between the sub- and supercritical regimes advanced by the
``standard'' model of magnetic support (e.g., \cite{Mousch76}; \cite{SAL87}).

\section{Discussion: A bigger question} \label{sec:discussion}

In the previous sections we have summarized results on the SFE in a
variety of contexts: driven vs.\ decaying simulations, and magnetic
versus non-magnetic. The main conclusions to be drawn from the existing
results are that (1) the very effect of the intensity of the turbulence
(measured by the rms Mach number $\Ms$) depends on whether the
turbulence is driven or decaying, and (2) the efficiency is reduced as
the magnetic field increases from zero to supercritical levels to
subcritical levels, but the values of the magnetic field strength
needed to attain realistic values of the SFE again depend on whether
the turbulence is driven of decaying. Thus, the behavior of the SFE with
the parameters $\Ms$ and $\mu$ is relatively well understood, but it is
necessary to determine what is the true nature of the turbulence driving
in molecular clouds (driven, decaying, or somewhere in between) in order
to assess the response of the SFE to the parameters.

It is also important to note that in all the simulations described
above, it is necessary to define a certain time at which to terminate
the accounting of the mass deposited in collapsed objects. This time is
typically a few to several free-fall times (a few Myr). If left to run
for arbitrarily long times, most of these simulations would eventually
turn most of their gas into stars. It is therefore also necessary to
address the MC lifetime problem in order to understand the SFE.
Presumably, accomplishing both tasks (determining the nature of the
driving and the clouds' lifetimes) amounts to addressing the questions
of how the clouds themselves form and acquire their properties, and how
they are eventually dispersed; that is, their full life cycle.

\section{Simulations of cloud formation and evolution} \label{sec:cloud_evol}

\subsection{Results} \label{sec:results}

The question of whether the turbulence is driven or decaying is
unsettled at present. Arguments in favor of continuous driving include
the fact that even nearly starless MCs such as Maddalena's cloud have
similar turbulent parameters as clouds with healthy star formation rates
(\cite{MT85}), and that CO clouds fall on a tight velocity
dispersion-size relation suggestive of a single cascade process
operating at scales ranging from $\sim 100$ pc to $\lesssim 0.1$ pc in
the ISM (e.g., \cite{Larson81}; \cite{HB04}; see also Breitschwerdt,
this volume). Also, the energy feedback from stellar sources once they
have started forming is thought to be able to possibly maintain the
turbulence in the clouds (e.g., \cite{Matzner02}; \cite{TKM06}; \cite{KMM06};
\cite{LN06}), or even disperse them altogether (e.g., \cite{FST94};
\cite{HBB01}; \cite{Ball04}). 

Recently, it has been proposed that MCs may acquire at least their
initial levels of turbulence from the very accumulation process that
forms the cloud (\cite{VBK03}; \cite{Heitsch_etal05}; \cite{VS_etal06a},
2006b; \cite{Heitsch_etal06}; see
also \cite{KI02}; \cite{IK04}), through a combination of the
thermal instability and various dynamical
instabilities in the compressed layer between converging flows. The
precise nature of the instability at work is not yet agreed upon.

These studies have shown that the collision of warm
neutral medium (WNM) streams at transonic velocities in the absence of
self-gravity produces velocity dispersions of several $\kms$, typical of
molecular clouds. Furthermore, \cite[\VS\ et al.\ (2006a)]{VS_etal06a}
also showed that the the pressure in the dense ($n> 100 \pcc$) gas is
larger than the mean WNM pressure by factors 1.5--5, due to the ram
pressure of the compressive motion that forms the clouds. These results
suggest that cloud formation by WNM stream collisions or passing shocks
can produce the observed turbulent velocity dispersions in MCs and at
least part of their excess pressure.

Most relevant for our discussion here are the facts that in those
studies the turbulence is driven for as long a time as the inflow that forms
the cloud persists, and that the rms Mach number of the turbulence in the dense
gas depends on the Mach number of the inflow (see also
\cite{FW06}). This means that, at least during the early epochs of a
molecular cloud's existence, the turbulence may be driven, albeit
presumably the driving rate itself is decaying, as the inflows that form
the cloud subside, and, eventually, the cloud may be left in a decaying
state.

This mechanism has been recently investigated including self-gravity and
a sink particle prescription for treating collapsed objects by
\cite[\VS\ et al.\ (2006b)]{VS_etal06b}. This study has shown that,
within its framework and limitations (magnetic fields, stellar energy
feedback and chemistry were not included), the clouds evolve secularly,
rather than achieving a quasi-stationary state.  The collision of WNM
streams nonlinearly triggers thermal instability and a transition to the
cold neutral medium. Due to the ram pressure of the inflows, densities
and temperatures overshoot to values typical of molecular gas. The dense
gas (the ``cloud'') evolves by continuing to incorporate mass,
generating an increasingly deep gravitational potential well in the process.
Eventually, the gravitational energy $\Eg$ of the cloud overwhelms the
thermal+turbulent energies ($\Eth$ and $\Ek$) and the cloud begins to
contract gravitationally. This process is illustrated in fig.\
\ref{fig:SFE_non-MHD} (\emph{right panel}), which shows the evolution of
the dense gas and stellar mass in the simulation, along with the various
energies for a simulation in a cubic box of 256 pc per side, in which a
cloud is formed by the collision of two oppositely-directed WNM streams
at speeds of $\pm 9.2~\kms$, and each with a length of 112 pc and a
radius of 32 pc.


In this simulation, $\Eg$ is seen to become dominant at $t \sim 12$ Myr,
but the kinetic energy is ``dragged along'' by the gravitational contraction,
with the result that there is near equipartition between the two
throughout the collapse, in agreement with observations.
After some delay (at $t \sim 17$ Myr for this simulation), local collapse
events begin to occur, and within three more Myr ($t \sim 20$ Myr),
$\sim 15$\% of the cloud's mass ($\sim 5000 M_\sun$) has been converted
to stars, at a mean rate $\sim 1.7 \times 10^{-3} M_\sun$ yr$^{-1}$. In
the simulation, this rate continues for another $\sim 5$ Myr (to $t \sim
25$ Myr), but already by $t \sim 20$ Myr, the mass that has been
converted to stars implies that enough OB stars should be present to
destroy the cloud (\cite{FST94}), assuming a standard IMF.  The SFE in
this simulation at this time ($\sim 15$\%) is thus comparable to that in 
the simulations of gravitationally bound clouds 
discussed in \S \ref{sec:Eff_M}. But, as in all those
simulations, this is dependent on the assumption that the cloud somehow
ceases to form stars some 3--5 Myr after it started. 

\subsection{Implications} \label{sec:implic}

Some important consequences of this scenario for MC formation should be
noted. First, even though a long delay ($\sim$ 15 Myr) occurs
between the beginning of the formation process (the time at which the
collision between the WNM streams begins), the cloud is expected to
remain atomic during most of this time, since the cloud's mean column
density is only reaching typical values for molecule formation ($\sim
10^{21} \psc \sim 8 M_\sun$ pc$^{-2}$; see \cite{FC86}; \cite{HBB01} and
references therein; Blitz, this volume) by the time it is beginning to
form stars. Thus, even though the cloud as a density enhancement lives
$\sim 20$ Myr, its \emph{molecular} stage is expected to comprise only the
last few Myr. That is, there may indeed be a long ``dormancy'' period
before the onset of star formation as suggested by various groups (e.g.,
\cite{PS00}, 2002; \cite{GL05}; \cite[Mouschovias et al.\ 2006]{MTK06}),
but most likely it is spent in an atomic, growing state, rather than in a
molecular, quasi-equilibrium one.

Second, this scenario of molecular cloud formation implies that the
mass-to-flux ratio of the cloud is a variable quantity as the cloud
evolves. This ratio is equivalent to the ratio of column density to
magnetic field strength (\cite{NN78}), with the critical colum density
given by $\Sigma \sim 1.5 \times 10^{21} \left[B/5\mu G \right]~\psc$.
Although in principle under ideal MHD conditions the criticality of a
magnetic flux tube involves \emph{all} of the mass contained within it, in
practice it is only the mass in the dense gas phase that matters,
because the diffuse gas is not significantly self-gravitating at the
size scales of MC complexes. As pointed out by \cite{HBB01}, the above
value of the dense gas' column density is very close to that required
for gravitational binding, and therefore, the cloud is expected to
become magnetically supercritical nearly at the same time it is becoming
molecular and self-gravitating. This is consistent with the results of
the simulations by \cite[\VS\ et al.\ (2006b)]{VS_etal06b}, in which the
column densities of the first four regions to form stars were measured
to have column densities within a factor of two of $ N = 10^{21} \psc$
immediately before the first local collapse event occurred there.

Finally, the results from the simulations by  \cite[\VS\ et al.\
(2006b)]{VS_etal06b} would seem to suggest a return to the
\cite[Goldreich \& Kwan (1974)]{GK74} scenario of global gravitational
collapse in MCs, except that the criticism by \cite[Zuckerman \& Palmer
(1974)]{ZP74} would be avoided in part because the nonlinear turbulent density
fluctuations collapse earlier than the whole cloud, involving only a
fraction of the total mass, and in part because as soon as the stars form
they probably contribute to dispersing the cloud, or at least halting
its global collapse. This is consistent with the recent suggestion
by \cite{HB06} that the Orion MC may be undergoing global gravitational
collapse.

\section{Conclusions} \label{sec:conclusions}

We conclude that numerical simulations of isolated clouds up to the
present have quantitatively constrained the effect of the turbulent Mach
number and the magnetic field strength on the SFE, but in turn this effect
depends on the nature of the turbulence production and maintenance, and
on the lifetimes of the clouds themselves. Simulations of MC formation
within their diffuse environment have begun to shed light on these
issues, but much parameter space exploration and inclusion of additional
physics (notably, magnetic fields and stellar energy feedback)
remain to be done.

\begin{acknowledgments}
The author gratefully thanks J. \BP\ for useful comments on the
man\-u\-script. This work has received financial support from CRyA-UNAM
and CONACYT grant
47366-F. The numerical simulation described in \S \ref{sec:results} was
performed on the linux cluster at CRyA-UNAM acquired with funds from
CONACYT grant 36571-E.
\end{acknowledgments}

\begin{discussion}
\discuss{Clark}{You say that the final SFE in your simulation is $\sim
15$\%, which is too high. Have you tried unbound clouds (kinetically) to
see whether the final SFE goes down?}

\discuss{\VS} {In this simulation the boundedness of the cloud is
produced self-consistently by the cloud-formation process, so we do not
control it directly. In this particular simulation, the resulting cloud
is strongly bound.  Nevertheless, one could try to produce a less
strongly bound cloud, or even unbound, by decreasing the mass contained
in the inflowing streams, or increasing their speed. We are currently
performing a parameter study to investigate different cloud masses and
inflow velocities.}

\discuss{Rosolowsky}{Could you comment on the applicability of your
simulations to the formation of GMCs, specifically in the case where the
scales over which you have to gather gas become significant on a
galactic scale?}

\discuss{\VS} {
I am convinced that the process of compression, then cooling with
turbulence generation, and finally gravitational collapse, should be
representative of GMC formation in spiral arms, although
modeling the process more accurately should incorporate the vertical
stratification as well.}

\end{discussion}

\end{document}